\documentclass{article}

\usepackage{arxiv}

\usepackage[utf8]{inputenc} 
\usepackage[T1]{fontenc}    
\usepackage{hyperref}       
\usepackage{url}            
\usepackage{booktabs}       
\usepackage{amsfonts}       
\usepackage{nicefrac}       
\usepackage{microtype}      
\usepackage{lipsum}
\usepackage{graphicx}
\graphicspath{ {./images/} }

\title{Passive Aperiodic Optical Phased Array based on Uniform Random Shuffle}

\author{
 Bowen Yu \\
  Dept. of Electrical and Computer Engineering\\
  University of Michigan\\
  Dearborn, MI 48128 \\
  \texttt{bowenyu@umich.edu} \\
   \And
 Dachuan Wu \\
  Dept. of Electrical and Computer Engineering\\
  University of Michigan\\
  Dearborn, MI 48128 \\
  \texttt{dachuanw@umich.edu} \\
  \And
 Yasha Yi \\
  Dept. of Electrical and Computer Engineering\\
  University of Michigan\\
  Dearborn, MI 48128 \\
  Energy Institute,\\
  University of Michigan\\
  Ann Arbor, MI 48109 \\
  \texttt{yashayi@umich.edu} \\
}

\begin{document}
\maketitle
\begin{abstract}
Grating lobes arise from the periodic nature of element spacing in the optical phased array. Essentially, the phased array performs the Spatial Fourier Transform on light; the steering capability of the main lobe is governed by phase shift variations among waveguides, and the Sidelobe Suppression Ratio (SLSR) correlates with the uniformity of emitter positions. Leveraging this understanding, we have optimized a 1×64 channel passive aperiodic OPAs with the uniform random shuffle in the emitter's position. Our conceptual simulations highlight a robust steering capability (18.60° / 10nm) and SLSR (-13.46 dB @ 0° / -8.27 dB @ ±45°), and initial measurements demonstrate the steering capability (9.8 ° / 10nm, with smaller phase shifts design) and SLSR (-6.1dB @ -33.4°) from the preliminary fabrication.
\end{abstract}


\section{Introduction} 
Most research on the optical phased array systems (OPAs) starts from the integrated approach from silicon on insulators (SOI) with periodically arranged emitters \cite{Van:09}. Over the years, numerous review papers have been published, offering various insights and feasibilities of OPAs and LiDAR technologies based on photonic integrated circuits \cite{Heck:17,Sun:19,Guo:21,Kim:21,Hsu:22,Wu:23}.  Some research efforts have been dedicated to system-level integration, employing simple periodic emitting patterns. A notable issue with this approach is the emergence of prominent grating lobes in the far field, attributed to the uniform spacing of the emitter arrays. In LiDAR applications, these lobes may be incorrectly identified as the main lobe, leading to potential inaccuracies. To address this challenge, other research in the field of OPAs has shifted focus to utilizing aperiodic emitter pitches, aiming to attain a broader steering range and higher resolution \cite{Zhuang:18,Yang:20,Du:22,Kwong:11,Hutchison:16,Fatemi:19,Kazemian:21,Xue:22,Fukui:21,Komljenovic:17,Wang:22}. This is done by optimizing the spacing between waveguides and expanding the emission aperture, thus preventing aliasing across the field of view and enabling wide-angle emissions from individual antennas. These advancements often employ mathematical and machine learning techniques like particle swarm optimization, genetic algorithms, NSGA-II, and pattern search methods for optimal emitter placement. These methods have proven highly effective; for instance, one study achieved -13.39dB @ 0° in a 1×64-channel OPAs \cite{Wang:22}. From the optimization perspective, it is crucial to manage the trade-off between sidelobe level and beam width, with computational tools and simulations being essential due to the complexity of these designs. Despite these advancements, the ultimate spatial resolution limits of non-uniform OPAs, particularly the theoretical limits of the Sidelobe Suppression Ratio (SLSR), remain unresolved. Essentially, phased arrays can be viewed as the diffraction model, performing a Spatial Fourier Transform on the light. Our previous work demonstrated a high-efficiency 3D passive wavelength-tuning 4×16 OPAs device on a SiN/SiO\textsubscript{2} platform, but it showed noticeable grating lobes due to uniform emitter spacing \cite{Wu:23}.

Leveraging the principles from the diffraction model, we optimized our $\Omega$-shape OPAs for high SLSR and steering capability in this study. We introduced a uniform random shuffle in emitters' positions to break the array's periodicity, and integrated large passive phase delay lines for increased phase delay variation. To optimize our simulation process, we developed the Rapid Farfield Approximation, a simplified structural model that significantly reduced computational efforts. The simulation results validated our method for achieving high SLSR and steering rates. Additionally, our preliminary fabrication and characterization, despite some fabrication errors and equipment limits, indicate the potential viability of these strategies for future OPAs designs.

\section{Design and simulation}
\subsection{Correlating diffraction model to 1D OPAs design}

Similar to the diffraction model, grating lobes in any optical phased array system are caused by the intrinsic periodicity among emitting elements. The Fourier transform effectively quantifies this periodicity within these arrays, offering a measure of its extent. The farfield E-field equation (Eq. \ref{eq:refname1}) \cite{Du:22} for any arbitrary phased arrays and the Discrete Fourier Transform (DFT) equation (Eq. \ref{eq:refname2}) are the keys to the analysis:
\begin{equation}
E(\theta)=\sum_{i=1}^{N}A_{i}e^{j(2\pi\frac{y_{i}}{\lambda}sin\theta)}e^{-j\phi_{i}}
\label{eq:refname1}
\end{equation}
\begin{equation}
F(k)=\sum_{n=0}^{N-1}f(n)e^{j(2\pi\frac{n}{N}k)}
\label{eq:refname2}
\end{equation}

In Eq. \ref{eq:refname1}, $A_{i}$ and $\phi_{i}$ are the amplitude and phase of each emitter, $y_{i}$ is the emitter position and $\theta$ is the viewing angle. For $f(n)$ in the DFT equation (Eq. \ref{eq:refname2}), its ordinal value $n$ signifies real space position, and its cardinal value $f(n)$ indicates the presence (1) or absence (0) of an emitter at   $n$. $k$ is the wave vector, equivalent to $1/\Lambda$, where $\Lambda$ is the periodicity of the array. For an aperiodic arrangement, $\Lambda$ is estimated as the arrays' size divided by the number of emitters. 

\begin{figure}[ht!]
\centering\includegraphics{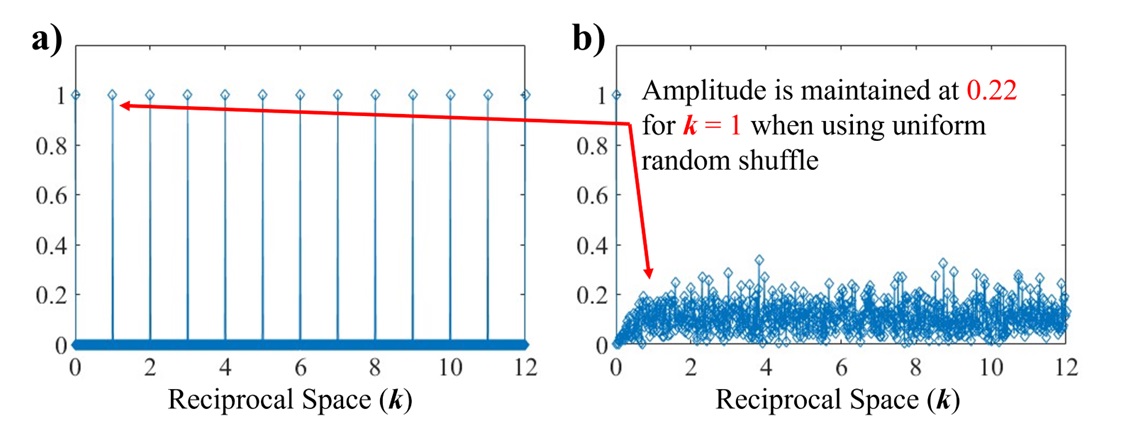}
\caption{Spatial Fourier Transform of \textbf{a)} periodic and \textbf{b)} aperiodic array arrangement (with uniform random shuffle) sets.}
\label{fig:1}
\end{figure}

The first exponential term in Eq. \ref{eq:refname1}, close to the exponential in Eq. \ref{eq:refname2}, performs a Spatial Fourier Transform on light from emitters. To achieve a high SLSR, we first employ sets of emitters' positional shifts via random uniform distribution to maximize disorderness in sample variations, as it possesses the highest entropy among different discrete distributions. Subsequently, we apply a uniform random shuffle, specifically the Fisher-Yates Shuffle, to the second set of positional shifts  (Fig. \ref{fig:1}). To ensure statistical consistency (mean and variance) throughout the random generation process, we generate shifting arrays with even distribution over a specified range, thereby minimizing crosstalk. 

The second exponential term in Eq. \ref{eq:refname1} represents the shift property within the Spatial Fourier Transform, derived from the Huygens model. In the simplicity of a single-mode waveguide-based design, we achieve phase delays through varying path lengths. The phase delay for each waveguide is mathematically expressed as:

\begin{equation}
\phi_{i} = \frac{2 \pi L_{i} n_{eff}(\lambda_{0})}{\lambda_{0}}
\label{eq:refname3}
\end{equation}

\subsection{Rapid Farfield Approximation}
\begin{figure}[ht!]
\centering\includegraphics{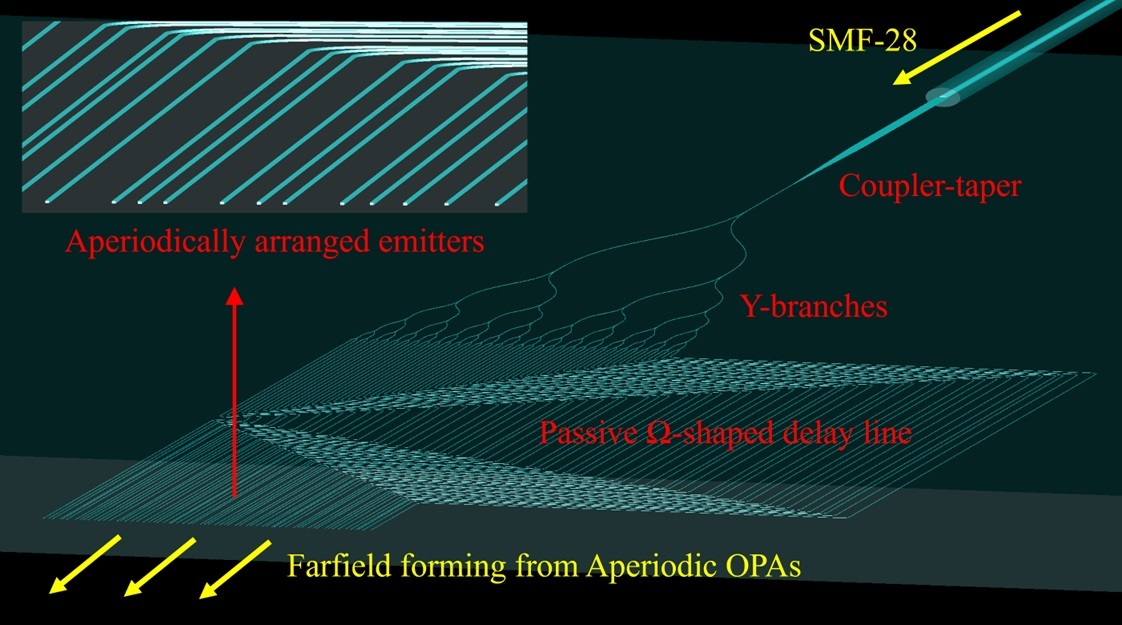}
\caption{Schematic of 1×64 passive aperiodic OPAs.}
\label{fig:2}
\end{figure}

In this study, expanding on our prior work \cite{Wu:23}, we evaluate the impact of aperiodic emitter layouts on steering and SLSR enhancement. Preferring simplicity, we chose a single-layer design over a multi-layer one, easing manufacturing while verifying concepts for future complex integrations (Fig. \ref{fig:2}). The design uses a frequency-tunable laser source, coupled to a waveguide via SMF-28 and an on-chip edge coupler-taper, supporting only TE\textsubscript{0} and TM\textsubscript{0} modes in 0.6 µm × 0.705 µm waveguides. We opt for the TM\textsubscript{0} mode due to its slightly better bend loss and minimal crosstalk spacing. The beam is then split into 1×64 channels, each with an $\Omega$-shaped delay line.

While the Spatial Fourier Transform offers a rough approximation of the farfield pattern, employing the Finite-Difference time domain (FDTD)  yields a more accurate representation at the cost of the computational workload. To simulate the farfield pattern efficiently using Lumerical's FDTD, we implement the Rapid Farfield Approximation (RFA) technique. This approach accounts for the linearly increasing delay line length in each waveguide as we move upwards from the bottom of the $\Omega$-shaped configuration. The passive phase delays are thus calculated as:

\begin{equation}
\Delta \phi = \frac{2 \pi \Delta L \ n_{eff}(\lambda_{0})}{\lambda_{0}}
\label{eq:refname4}
\end{equation}

\begin{equation}
\Delta L = \alpha Y + \beta
\label{eq:refname5}
\end{equation}

\begin{equation}
\beta=\frac{[\alpha Y n_{eff}(1.55\mu m)/1.55\mu m]\%\pi}{\pi} \times \frac{1.55\mu m}{n_{eff}(1.55\mu m)}
\label{eq:refname6}
\end{equation}

In these equations, $\Delta \phi$ is the phase delay in each waveguide, $\lambda_{0}$ is the source wavelength and $n_{eff}$ is the effective index for TM\textsubscript{0} at different  $\lambda_{0}$. $\Delta L$ represents the delay line length, with $\alpha$ being the path difference multiplier and $Y$ the pitch position of the emitters. The modulus operator $\%$ is used to calculate $\beta$, which serves as a length correction term to center the main lobe at 0° for $\lambda$\textsubscript{0}=1.55µm. The simplicity of our structure allows the FDTD solver to focus solely on simulating the electromagnetic field around the dielectric-air interface at the emitter’s end. This targeted approach significantly reduces the time required for farfield formation simulations, making large-scale and quantitative analysis feasible on our computer. 

\begin{figure}[ht!]
\centering\includegraphics{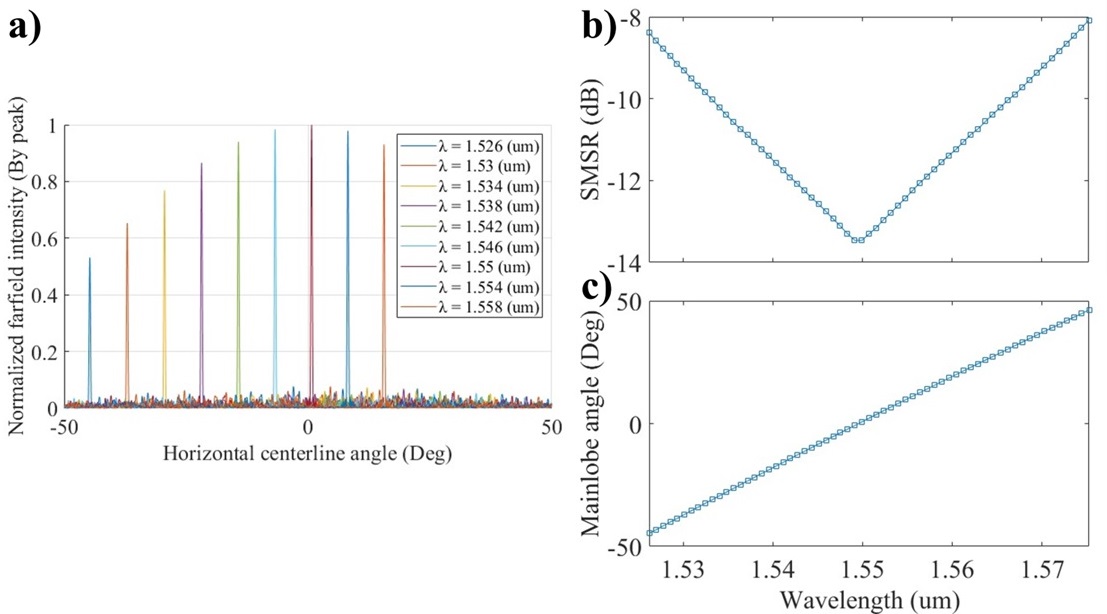}
\caption{\textbf{a)} RFA simulated farfield pattern at different $\lambda_{0}$. \textbf{b)} The corresponding SLSR and \textbf{c)} the mainlobe angles derived from these simulations.}
\label{fig:3}
\end{figure}

In 2D simulations, we applied 64 positional shifts to each waveguide from 6 µm periodic arrangement using a random uniform process, limiting maximum shifts to ± 2.4 µm to avoid crosstalk. The path difference multiplier ($\alpha$) was set to 16. We generated and analyzed 3000 sets, observing that SLSR decreased correspondingly with increased spatial frequency amplitude from 0.18 to 0.5, at $\xi$ = 0.1667 µm\textsuperscript{-1} or $\Lambda$ = 6 µm. To suppress the amplitude at $\Lambda$ = 6 µm, a second batch was processed using uniform random shuffle, effectively maintaining the amplitude to 0.22 (Fig. \ref{fig:1}b) and achieving an average SLSR of approximately -11.5 dB at 0°. We then proceeded to evaluate the performance of the best set from the simulations at different wavelengths, finding the peak SLSR to be -13.46 dB at 0° and -8.27 dB at ±45° (Fig. \ref{fig:3}). All simulations consistently showed a steering capability of 18.6°/10nm, influenced by the fixed path difference multiplier rather than the array arrangement. Additionally, the Full Width at Half Maximum (FWHM) of the main lobes was consistently about 0.3° in all simulations, mainly determined by the number of emitters. 3D simulations with a reduced path difference multiplier ($\alpha$ = 10) were conducted to anticipate fabrication variances. We tested periodic and aperiodic OPAs, which achieved -10.0 dB SLSR (aperiodic set) using 3D RFA. Utilizing Lumerical’s Eigenmode Expansion, we optimized the coupler-taper for -7.30 dB coupling efficiency from the SMF-28 fiber, and the y-branch splitter showed minimal loss (-0.534 dB per split), enhancing OPAs performance.

\section{Fabrication and experimentation}
\begin{figure}[ht!]
\centering\includegraphics{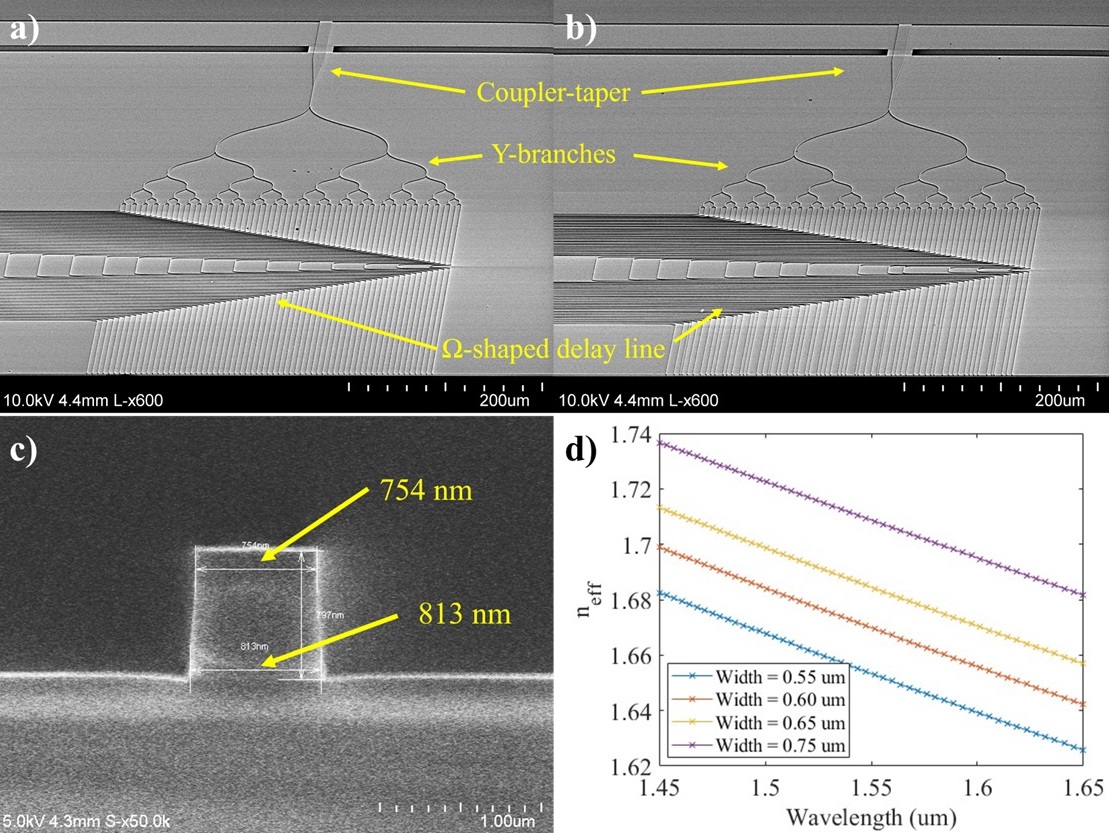}
\caption{\textbf{a \& b)} A tilted top view of \textbf{a)} periodic and \textbf{b)} aperiodic OPAs from the SEM. \textbf{c)} The cross-sectional view of a single waveguide from the aperiodic OPAs, offering detailed structural insights. \textbf{d)} The simulated effective index across various wavelengths for waveguides with differing widths.}
\label{fig:4}
\end{figure}

In the preliminary fabrication phase, we produced the two samples discussed in the previous section with the conventional SOI fabrication techniques \cite{Wu:24} from the Lurie Nanofabrication Facility (LNF). Overall, the structures showed all the necessary features of the design. However, we observed a discrepancy in waveguide width: measurements showed the waveguide has a width of 0.7 – 0.8 µm in trapezoid shape, instead of 0.6 µm width in rectangle shape which is used in our simulations, as illustrated in Fig. \ref{fig:4}a – c.

\begin{figure}[ht!]
\centering\includegraphics{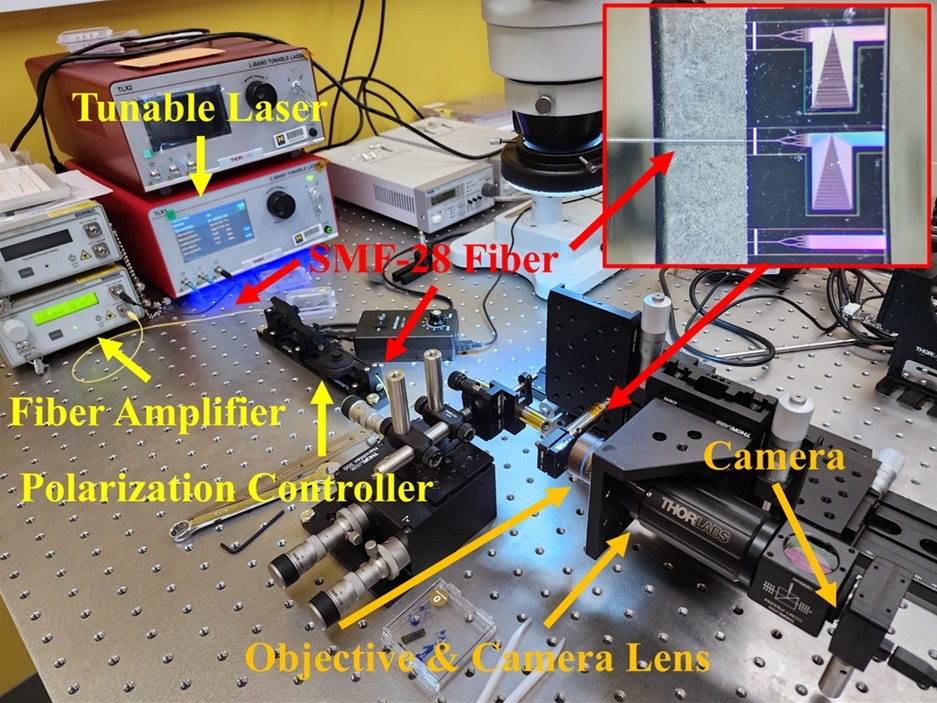}
\caption{Schematic of characterization setup.}
\label{fig:7}
\end{figure}

\begin{figure}
\centering\includegraphics{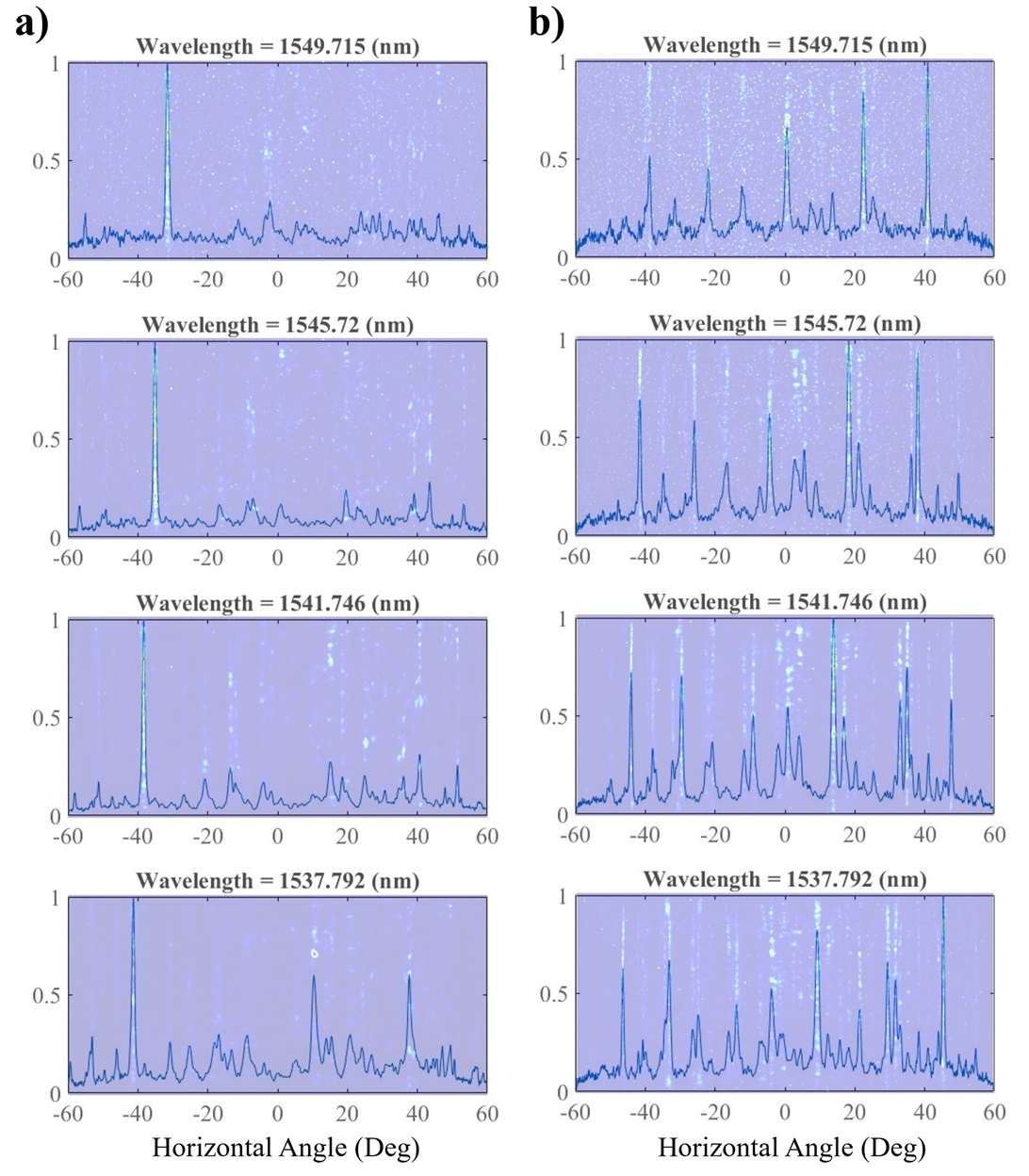}
\caption{Experimental measurement result for \textbf{a)} Aperodic OPAs and \textbf{b)} Periodic OPAs, each evaluated at different $\lambda$.}
\label{fig:5}
\end{figure}

To assess the steering capability, we implemented a simple 4f optical system comprising an objective and a camera lens (Fig. \ref{fig:7}). This setup enabled us to calculate farfield angles through trigonometric analysis based on the stage's position. We conducted farfield measurements across wavelengths from 1538 to 1550 nm, constrained by the sensitivity range of our camera (Point Grey, CMLN-13S2M). The results indicated a steering capability of 9.8°/10nm for the aperiodic OPAs and 11.3°/10nm for the periodic OPAs (Fig. \ref{fig:5}).  These observed values slightly differ from the simulated rate of 12.1°/10nm. On examining the grating lobes of periodic OPAs at various angles and wavelengths, we observed steering rates fluctuating between 11.3°/10nm near 0° and 9.7°/10nm around 40°. This variation is attributed to potential imaging distortions inherent in our basic 4f optical system, coupled with the resolution limitations of our 3-axis stages, which have a 50 µm resolution. This resolution limitation could introduce significant errors in angle estimation due to the typical lens positional shifts being within 200 µm. Alternatively, we employed a method of correlating simulated grating angles with actual grating angles and shifts in pixels from the images, leading to an estimated steering rate of 11.90°/10 nm for the periodic OPAs. 

\begin{figure}[ht!]
\centering\includegraphics{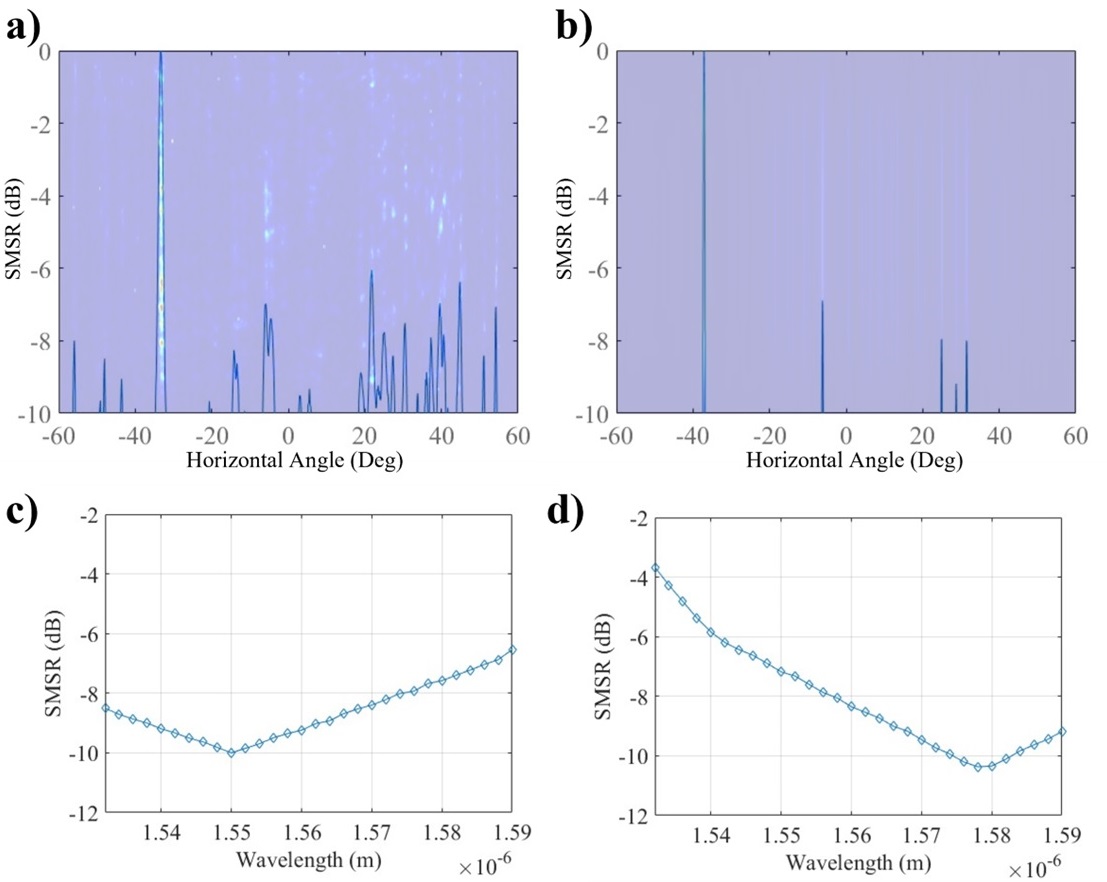}
\caption{\textbf{a)} The measured farfield pattern for aperiodic OPAs at $\lambda$ = 1.58 µm. \textbf{b)} The corresponding simulated farfield for waveguides at width = 0.75 µm. \textbf{c)} Illustrates the simulated SLSR for waveguides with a narrower width of 0.60 µm, while \textbf{d)} displays the simulated SLSR for the 0.75 µm waveguide width.}
\label{fig:6}
\end{figure}

To determine the SLSR, we tuned our laser (Thorlabs, TLX-1) to 1547.7 nm, a wavelength where our camera exhibits peak sensitivity. The farfield pattern analysis (Fig. \ref{fig:6}a) revealed that the SLSR of the aperiodic OPAs was -6.1 dB, markedly lower than expected value (-10.0 dB). The mainlobe is at -33.4° from the experiment and 41.2° from the simulation. A significant contributing factor is the shift in the effective index due to variations in waveguide, where an increase of 3\%  changes (Fig. \ref{fig:4}d). Subsequent simulations, accounting for this modified waveguide width (0.75 µm, rectangular), corroborated this effect, showing a decrease in SLSR to -6.6 dB (Fig. \ref{fig:6}b). \textbf{}This change also induced wavelength range shifts, with the 0° main lobe's wavelength moving from 1.55 µm to 1.58 µm (SLSR = -10.1dB @ $\lambda_{0}$=1.58 µm, Fig. \ref{fig:6}c-d). Additionally, we noticed an uneven intensity distribution in the nearfield pattern, likely resulting from variable losses across different waveguides, further contributing to the diminished SLSR from farfield converging.

\section{Conclusion \& Discussion}

Based on the understanding from the Spatial Fourier Transform, we developed a simple yet effective method to reduce grating lobes and increase the steering capability in a 1×64 channel passive aperiodic OPAs. We also implement RFA, a simplified model in farfield forming from the FDTD, to effectively conduct large-scale and quantity simulations. From this study, the effectiveness of the SLSR is primarily attributed to reducing the amplitude at specific spatial frequencies, especially at $\Lambda$. This is accomplished by incorporating uniform random permutated positional shifts to the emitter’s position. On the other hand, the steering capability is directly tied to the passive path differences in the $\Omega$-shaped waveguides. Greater passive path differences between waveguides lead to larger steering capabilities, though this comes with a trade-off in terms of lower engineering tolerance.

From the simulations, we demonstrate this aperiodic OPAs exhibiting the SLSR of -13.46 dB at 0° and -8.27 dB at ±45°, while the steering rate is 18.60°/10nm ($\alpha$ = 16). However, due to fabrication errors and characterization tools’ limits, the experimental SLSR is -6.1 dB at -33.4°, and the steering rate is 9.8°/10nm ($\alpha$ = 10). With the experience gained from the LNF, we believe that SLSR will be further improved by reducing the fabrication error. 

We acknowledge the potential for further optimization in our study. Our approach, which introduces and visualizes the disorder into the emitter arrangement to reduce sidelobes, draws upon principles from diffraction theory and related research. Future improvements might include exploring different emitter numbers and pitch sizes, as well as other possibilities like hyperuniform distributions, known for reducing density fluctuations at larger scales. Moreover, the concepts used in this study for 2D array arrangements are validated, suggesting the potential for optimizing 3D OPAs structures using similar strategies.

\bibliographystyle{unsrt}  
\bibliography{references}

\end{document}